\documentclass[12pt,preprint]{emulateapj}

\usepackage{graphicx,amsmath,natbib}
\usepackage{color}
\newcommand{\bd}{BD+45$^{\circ}$\,598}
\newcommand{\um}{\,$\mu$m}
\newcommand{\vrad}{$v_R$}

\newcommand{\loglxlbol}{log($L_{\rm X}$/$L_{bol}$)}

\slugcomment{Last Compiled \today}

\shorttitle{Discovery of a Circumstellar Debris Disk Around the $\beta$ Pic Moving Group Member \bd}
\shortauthors{Hinkley et al.}

\begin{document}

\title{Discovery of an Edge-on Circumstellar Debris Disk Around BD+45$^{\circ}$\,598: a Newly Identifed Member of the $\beta$ Pictoris Moving Group}

\author{Sasha Hinkley\altaffilmark{1}}
\author{Elisabeth C.~Matthews\altaffilmark{2}}
\author{Charl\`ene Lefevre\altaffilmark{3}}
\author{Jean-Francois Lestrade \altaffilmark{4}}
\author{Grant Kennedy\altaffilmark{5}}
\author{Dimitri Mawet\altaffilmark{6,7}}
\author{Karl R. Stapelfeldt\altaffilmark{7}}
\author{Shrishmoy Ray\altaffilmark{1}}
\author{Eric Mamajek\altaffilmark{7}}
\author{Brendan P. Bowler\altaffilmark{8}}
\author{David Wilner\altaffilmark{9}}
\author{Jonathan Williams\altaffilmark{10}}
\author{Megan Ansdell\altaffilmark{10,11}} 
\author{Mark Wyatt\altaffilmark {12}} 
\author{Alexis Lau\altaffilmark {13}}                  
\author{Mark W.~Phillips\altaffilmark{1}}       %Change to Hawaii? 
\author{Jorge Fernandez Fernandez\altaffilmark{5}}
\author{Jonathan Gagn\'e\altaffilmark{14,15}}
\author{Emma Bubb\altaffilmark{16}}
\author{Ben J. Sutlieff\altaffilmark{17,18}}
\author{Thomas J.~G.~Wilson\altaffilmark{1}}
\author{Brenda Matthews\altaffilmark{19,20}}
\author{Henry Ngo\altaffilmark{19}} %Herzberg
\author{Danielle Piskorz\altaffilmark{7}} %Need correct affiliation here. 
\author{Justin R. Crepp\altaffilmark{21}}
\author{Erica Gonzalez\altaffilmark{21}}
\author{Andrew W. Mann\altaffilmark{22}}
\author{Gregory Mace\altaffilmark{8}}
%\author{Adam Kraus?}
%\author{Ben Montet}
%\author{Ben Zuckerman?}

\altaffiltext{1}{University of Exeter, Astrophysics Group, Physics Building, Stocker Road, Exeter, EX4 4QL, United Kingdom }
\altaffiltext{2}{l'Observatoire Astronomique de l’Universit\'e de Gen\`eve.} 
\altaffiltext{3}{Institut de RadioAstronomie Millim\'etrique (IRAM), 300 rue de la Piscine, 38406 Saint Martin d’H\`eres, France}
\altaffiltext{4}{LERMA, Observatoire de Paris, PSL Research University, CNRS, Sorbonne Universit\'es, UPMC Univ.~Paris 06, 75014 Paris, France}
\altaffiltext{5}{Department of Physics, University of Warwick, Coventry CV4 7AL, United Kingdom}
\altaffiltext{6}{Department of Astronomy, California Institute of Technology, Mail Code 249-17, 1200 E. California Blvd, Pasadena, CA 91125}
\altaffiltext{7}{Jet Propulsion Laboratory, California Institute of Technology, M/S 321-100, 4800 Oak Grove Drive, Pasadena, CA 91109}
\altaffiltext{8}{Department of Astronomy, The University of Texas at Austin, 2515 Speedway Blvd. Stop C1400, Austin, TX 78712}
\altaffiltext{9}{Center for Astrophysics $\vert$ Harvard \& Smithsonian, 60 Garden St., MS 42, Cambridge, MA 02138}
\altaffiltext{10}{Institute for Astronomy, 2680 Woodlawn Drive, Honolulu, HI 96822-1897}
\altaffiltext{11}{NASA Headquarters, 300 E St. SW, Washington DC  20546}
\altaffiltext{12}{Institute of Astronomy, University of Cambridge, Madingley Road, Cambridge, CB3 0HA, United Kingdom}
\altaffiltext{13}{Aix Marseille Univ, CNRS, CNES, LAM, Marseille, France}
\altaffiltext{14}{Plan\'etarium Rio Tinto Alcan, Espace pour la vie, 4801 av. Pierre-De Coubertin, Montr\'eal, QC H1V~3V4, Canada}
\altaffiltext{15}{Institute for Research on Exoplanets, Universit\'e de Montr\'eal, D\'epartement de Physique, C.P.~6128 Succ. Centre-ville, Montr\'eal, QC H3C~3J7, Canada}
\altaffiltext{16}{Institute for Astronomy, The University of Edinburgh, Royal Observatory, Blackford Hill, Edinburgh EH9 3HJ UK}
\altaffiltext{17}{Anton Pannekoek Institute for Astronomy, University of Amsterdam, Science Park 904, 1098 XH Amsterdam, The Netherlands}
\altaffiltext{18}{Leiden Observatory, Leiden University, P.O. Box 9513, 2300 RA Leiden, The Netherlands}
\altaffiltext{19}{Herzberg Astronomy \& Astrophysics Research Centre, National Research Council of Canada, 5071 West Saanich Road, Victoria, BC, V9E 2E7 Canada.}
\altaffiltext{20}{Department of Physics \& Astronomy, University of Victoria, 3800 Finnerty Road, Victoria, BC, V8P 5C2 Canada}
\altaffiltext{21}{Department of Physics, University of Notre Dame, 225 Nieuwland Science Hall, Notre Dame, IN 46556}
\altaffiltext{22}{University of North Carolina, Department of Physics \& Astronomy, 120 E. Cameron Ave., Phillips Hall CB3255, Chapel Hill, NC 27599}

\begin{abstract}
We report the discovery of a circumstellar debris disk viewed nearly edge-on  and associated with the young, K1 star \bd~using high-contrast imaging at 2.2\,$\mu$m obtained at the W.M.~Keck Observatory. 
We detect the disk in scattered light with a peak significance of $\sim$5$\sigma$ over three epochs, and our best-fit model of the disk is an almost edge-on $\sim$70\,AU ring, with inclination angle $\sim$87$^\circ$.  
Using the NOEMA interferometer at the Plateau de Bure Observatory operating at 1.3\,mm, we find resolved continuum emission aligned with the ring structure seen in the 2.2\um~images. 
We estimate a fractional infrared luminosity of $L_{IR}/L_{tot}$ $\simeq6^{+2}_{-1}$\,$\times$\,$10^{-4}$, higher than that of the debris disk around AU Mic. 
Several characteristics of \bd, such as its galactic space motion, placement in a color-magnitude diagram, and strong presence of Lithium, are all consistent with its membership in the $\beta$ Pictoris Moving Group with an age of 23\,$\pm$\,3 Myr. 
However, the galactic position for \bd~is slightly discrepant from previously-known members of the $\beta$ Pictoris Moving Group, possibly indicating an extension of members of this moving group to distances of at least 70\,pc. 
\bd~appears to be an example from a population of young circumstellar debris systems associated with newly identified members of young moving groups that can be imaged in scattered light, key objects for mapping out the early evolution of planetary systems from $\sim$10-100 Myr. 
This target will also be ideal for northern-hemisphere, high-contrast imaging platforms to search for self-luminous, planetary mass companions residing in this system.  
\end{abstract}


\section{Introduction}

\begin{figure*}[ht]
  \centering
  \vspace{-0.75in}
  \includegraphics[width=1.05\textwidth]{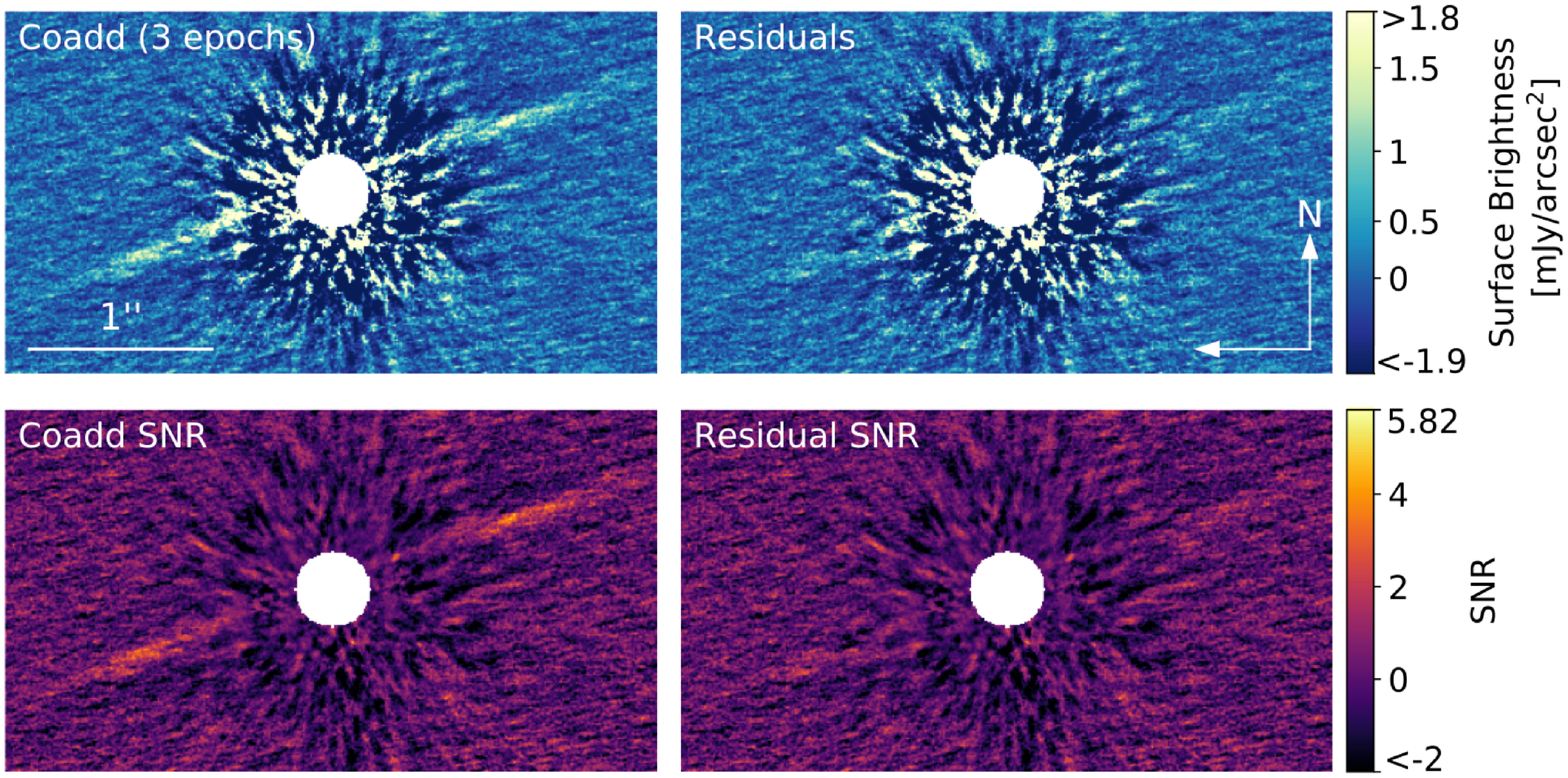}
  \vspace{-0.75in}
  %\caption{\textit{Left:} An image showing a co-addition of the  2.2\um~data from the three epochs of Keck/NIRC2 data. \textit{Right:} The co-added residual image of all three epochs after the best-fit model has been subtracted. The weighted average of the three models presented in Table~\ref{tab:fittable} has been subtracted from the cleaned data, which is then processed using the KLIP algorithm. No significant structure is seen in the residuals, indicating that our model is a good fit to the disk, to the limit of the signal-to-noise of these observations.}
  \caption{
  \textit{Top row:} Co-added 2.2\,$\mu$m data for the three high-quality epochs (\textit{top left}), and the co-added residual map with the best-fitting model subtracted from each epoch (\textit{top right}).  The weighted average of the three models presented in Table \ref{obstable} has been subtracted from the cleaned data, which is then processed using the KLIP algorithm. While some of the stellar speckles are saturated in this image, the full dynamic range of the disk signal is displayed. \textit{Bottom row:} Signal-to-noise maps of the coadded image and the residual image. The disk is detected at lower significance at very small separations from the star, due to the higher level of residual scattered starlight, as well as  ``self-subtraction'' due to the KLIP algorithm. In the residual images, a faint signal can be seen along the plane of the disk at low significance. %Future deep observations could further constrain the structure of this disk.
  The top images are scaled with the square of the intensity, while the bottom images are scaled linearly.
  }
  \label{fig:coadded}
\end{figure*}

Circumstellar debris disks are composed of the rocky or icy planetesimal bodies created in the planet formation process, as well as the tenuous dust and gas generated by their ongoing collisions \citep[e.g.][]{w08, mkw14, hdm18}.
The dust in debris disk systems, the presence of which is usually detected by excess far-infrared emission above the stellar photosphere \citep[e.g.][]{agb84}, is inherently short-lived, as the small released particles should be removed either by radiative forces, or accretion inwards by Poynting-Robertson drag \citep{bp91, dd03,kb04}.  
The long-term presence of dust in these systems thus suggests that it is being replenished by the continuous fragmentation of the parent planetesimals, which have been excited to eccentric orbits, causing them to experience ongoing collisions  \citep[e.g.][]{wd02,rss05}. This planetesimal fragmentation can be driven by several physical mechanisms, including perturbations by the most massive planetesimals \citep{kb08,ps12}, or a planetary mass companion embedded within or exterior to the disk \citep{kb04, w05, mw09}.
Thus, in addition to identifying themselves as a site of efficient planetesimal formation, stars with large amounts of circumstellar dust \textit{may} may be more likely to host massive planets at wide orbital separations \citep[e.g.][]{mlp97, anl01} suitable for direct imaging. However, no significant correlation between debris disk brightness and planets identified by any discovery method has yet been found \citep{bryden09, mmk15, mmb17, yg18}.

Surveys carried out by space-based observatories such as \textit{Spitzer}, \textit{Herschel} and \textit{WISE} have uncovered hundreds of stars within $\sim$200\,pc with a significant infrared excess in their spectral energy distributions at 22-24\um~suggesting the presence of warm circumstellar dust \citep[e.g.][]{tbb08, cbm09, cs16, pmh17, silverberg18}, 
%(Rieke 2005, Smith 2006, Rebull 2008, Trilling et al 2008, Carpenter 2009, Moor 2009), 
or a colder component identified by an excess at 70-160\um\,\citep[e.g.][]{hck08, msk10, emm13, skw18}.  
The identification of these circumstellar disks has also coincided with several high contrast imaging surveys targeting stars with an infrared excess \citep[e.g.][]{rcl13,jbm13,ndm19,lhq20}.  
Scattered-light images of debris disk systems in the optical or near-infrared ($\lesssim$2\,$\mu$m) returned by these surveys \citep[e.g.][]{ekf20} are particularly powerful for characterising planetary systems, as the morphology of stable debris belts may reveal the gravitational influence of giant planets sculpting the dust distribution through resonant interactions, \citep[e.g.][]{w03, ckk09, mrw11,srm13,mhv18}. 
When combined with a flux density distribution (or spectral energy distribution, ``SED'') of the system with adequate wavelength coverage, the spatial distribution of the dust structure can also provide constraints on the dust grain composition and size distribution.

To characterise planetary systems at the earliest stages, and specifically to identify where planets reside shortly after the epoch of planet formation has ended, the \textit{youngest} debris systems are particularly important.     
The nearest OB Associations \citep[e.g.~the Scorpius-Centaurus Association, $\sim$10-20\,Myr,][]{pmb12} have provided numerous scattered light images of young debris disks \citep[e.g.][]{mhv17,bonnefoy17, ekf20}. However, these associations are inherently at greater distances (~$\sim$140\,pc),  limiting the achievable physical resolution of 8-10m telescopes.    
The best compromise offering close proximity ($\sim$20-60\,pc) and relative youth ($\sim$10-100 Myr) are therefore nearby Young Moving Groups \citep[``YMGs'',][]{tqd06, bmn15, gmm18}: loose associations of gravitationally unbound stars sharing similar galactic space motions. 
Indeed, early discoveries of debris disks around YMG members have been extremely important for the development of our understanding of planetary systems  \citep{st84, rsw08, drc12, rbm14, mka16}.
The later-type members (spectral types K and later)  of YMGs should form the majority of the population of these groups, but have only begun to be identified in large numbers in the last several years, partly with the arrival of kinematic data from the \textit{Gaia} mission \citep[e.g.][]{ksa14,gf18,gbv18,bhz19}.     
However, $\sim$90\% of all debris disks\footnote{Data obtained from webdisks.jpl.nasa.gov, accessed on 18 May 2020} imaged in scattered light are associated with spectral types earlier than K, due to observational limitations, or the higher surface brightness of disks associated with early-type stars that have higher luminosity.
Therefore, late-type members of nearby YMGs may represent the best opportunity to identify a yet undiscovered population of young debris disks, key pieces in our process of mapping out the evolution of planetary systems from $\sim$10-100 Myr.

\bd~(2MASS J02211307+4600070, TYC 3294-2222-1) is a young, active, Li-rich K1-type star at 73.180\,$\pm$\,0.123 pc \citep{GaiaEDR3}
\footnote{Distance calculated as $D$ = 1/$\varpi$ with $\varpi$ = 13.6480\,$\pm$\,0.0230 mas from \citet{GaiaEDR3}, corrected by -0.017\,mas following \citet{Lindegren20}.}
that was first characterized in a spectroscopic survey of ROSAT All-Sky Survey (RASS) X-ray sources by \citet{gkf09}, and later flagged by \citet{mpk11} as a possible member of the $\beta$ Pictoris Moving Group (BPMG, hereafter). Based on its 22\,$\mu$m excess, \bd~was also identified in \citet{cs16} as one of their ``reserve'' IR excess stars, reporting a $L_{IR}/L_{star}$=3.9$\times$10$^{-4}$, and a dust temperature of 240\,K.  Note however that \cite{cs16} report an excess only in the 22$\mu$m band, and the temperature and luminosity of a two-parameter blackbody cannot be constrained by a single data point.  The WISE emission could therefore arise from a more luminous dust population that resides much farther from the star and at a much cooler temperature.

In this paper, we report the discovery of an edge-on debris ring associated with \bd~obtained using high contrast imaging at the W.M.~Keck Observatory (Figure~\ref{fig:coadded}). The clear signatures of youth and K1 spectral type make it a valuable object for the characterization of the early evolution of planetary systems surrounding late-type stars, similar to our own Solar System.  
In  \S\ref{sec:observations} we give an overview of our near-IR and millimeter observations and data processing, and in \S\ref{sec:modeling} we describe our model fitting process to both the infrared scattered light and millimeter images of the ring. 
In \S\ref{sec:age}, we show that this object is a likely member of the BPMG, and we summarize in \S\ref{sec:conclusions}.

\section{Observations \& Data Post Processing}\label{sec:observations}
Below we describe our observational coverage of \bd~at the W.M.~Keck Observatory, the James Clerk Maxwell Telescope (JCMT), Submillimeter Array (SMA), the Institut de RadioAstronomie Millim\'etrique (IRAM), and the Immersion Grating INfrared Spectrometer (IGRINS) at McDonald Observatory.

\begin{figure*}[ht]
  \centering
  \includegraphics[width=1.0\textwidth]{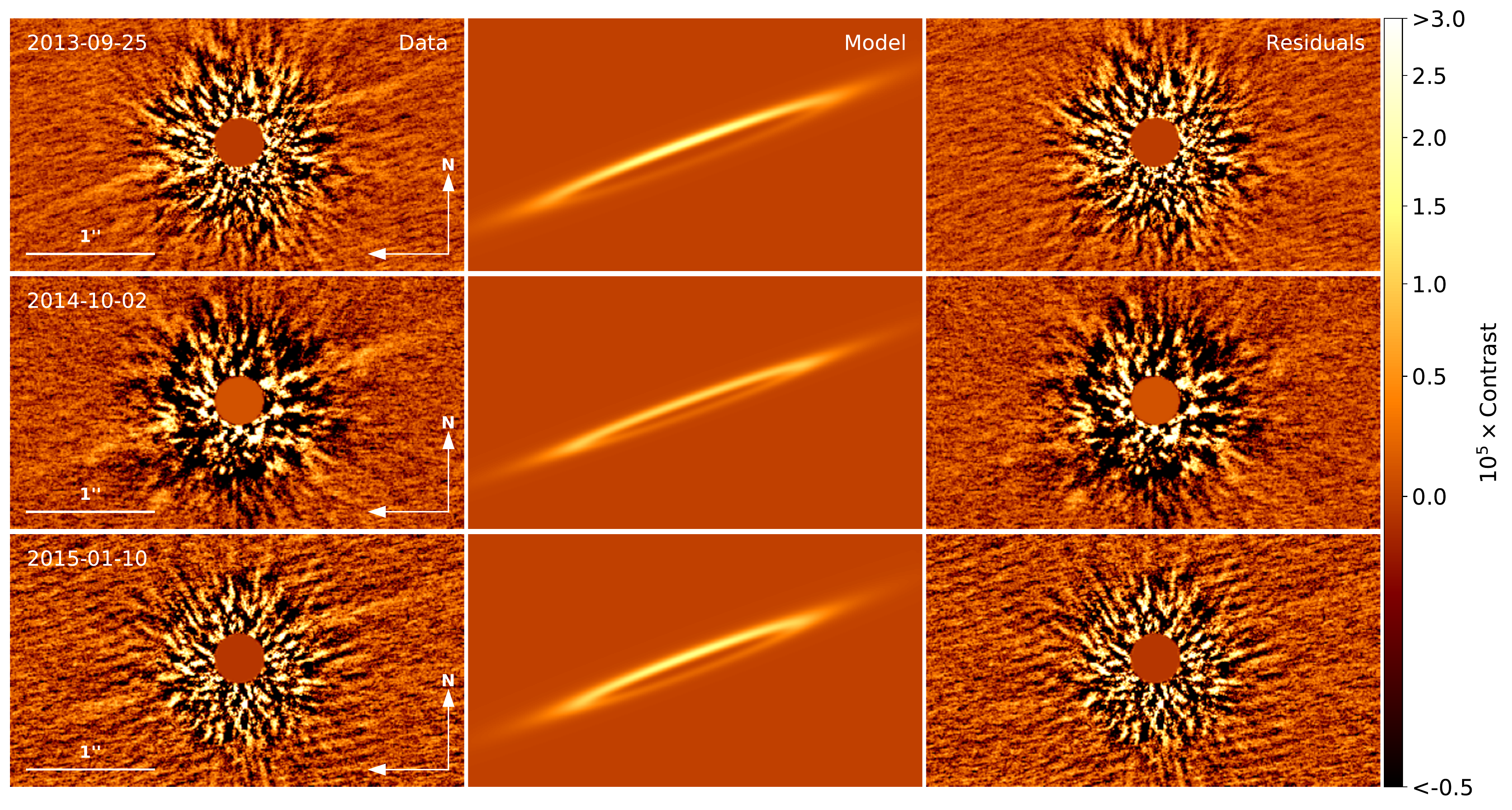}
  \caption{
 Processed 2.2\um~data (left column) and best fitting models (middle) with corresponding residuals (right) for the three Keck/NIRC2 epochs in which the debris disk was detected.  Full details of the fitting process and derived values are included in Section \ref{sec:modeling}. }
  \label{fig:three panel}
\end{figure*}

\subsection{Scattered Light Observations at the W.M.~Keck Observatory}\label{sec:keck_observations}
 Table~\ref{obstable} provides the detailed parameters of our observations of \bd, which was first observed on UT 2013 September 25 with the NIRC2 instrument at the W.M.~Keck Observatory using the $K^{\prime}$ filter (2.124\,$\mu$m central wavelength) and the 400 milliarcsecond coronagraphic mask.  The two subsequent epochs of observation of \bd~were obtained using this mask, and the 200 milliarcsecond radius of this mask effectively sets the ultimate inner working angle for our observations.  Seeing measurements obtained at Keck indicate that the average 0.5$\mu$m seeing for the 2013 and 2014 epochs ranged from 0.5-0.6$^{\prime\prime}$, but was poorer ($\sim$0.8$^{\prime\prime}$) for the 2015 epoch. All scattered light observations were obtained using the Keck Adaptive Optics (``AO'') system \citep{was00}, operating in pupil tracking mode \citep{mld06}, to differentiate quasi-static speckles \citep[e.g.][]{hos07} from any bona fide astrophysical sources. To remove the residual, uncorrected starlight, we model this speckle pattern using a customized algorithm based on projections of the data onto Karhunen-Lo\`eve eigenimages \citep[the Karhunen-Lo\`eve Image Processing or ``KLIP'' algorithm,][]{spl12}. Subtracting this model from each image provides deep contrast relative to the host star.  However, because of the $\sim$46$^\circ$ declination of \bd, Table~\ref{obstable} shows that typically only 18$^\circ$-36$^\circ$ of parallactic angle rotation could be achieved for each of our three epochs of scattered light imaging, meaning a substantial amount of residual scattered starlight contamination remains within $\sim$0.5$^{\prime\prime}$.  
 
 In Figure~\ref{fig:three panel} we show an image of \bd~with our three epochs of data coadded together, as well as a corresponding signal-to-noise (SNR) map of the same image.  The disk was immediately identified in our first epoch (Figure~\ref{fig:three panel}, \textit{top panels}), extending to $\sim$1\arcsec on each side of the star.  The disk structure in the images was robust against the number of eigenmodes chosen in the post-processing algorithm, further indicating that the structure is a true astrophysical signal, and not residual, scattered starlight.  However, to verify that the signal was indeed a physical structure associated with the star, \bd~was again detected using the $K^{\prime}$ filter at two subsequent epochs, 2014 October 02 and 2015 January 10 (see Figure~\ref{fig:three panel}). We also attempted another observation of \bd~on 2014 November 11, but this dataset suffered from poor atmospheric conditions corresponding to poor Adaptive Optics correction, resulting in a non-detection of the disk.

\subsection{JCMT \& SMA }
\bd~was not observed by either the \textit{Spitzer} or \textit{Herschel} observatories, and there are only upper limits available at 60\,$\mu$m and 100\,$\mu$m from \textit{IRAS}. So, as an additional check that this structure seen in the Keck images was not an artifact from the image post-processing, \bd~was observed at various dates betwen 2016 February 19 and 2016 August 10 for a total of  4.5 hours at 850\,$\mu$m (Band 3) using the SCUBA-2 instrument at the James Clerk Maxwell Telescope. The SCUBA-2 observations achieved an RMS sensitivity of 1.1 mJy per 14.5 arcsecond beam, but no signal at the location of \bd~was detected. These observations thus provide an upper limit of 3.3 mJy at 850\,$\mu$m, and we do not include these limits for our analysis.
Using the Submillimeter Array (SMA) on 2015 August 10, we observed \bd~at 224 GHz (1.3mm) in a six hour ``filler track.''  The array configuration using six antennas resulted in a $5.0\times2.7^{\prime\prime}$ beam, and using a point-source fit to the imaged visibilities, our observations revealed a $\sim$3.3$\sigma$ detection of a source with a flux density of 2.17 $\pm$ 0.76 mJy,  coincident with the location of the disk seen in scattered light. The observations followed standard procedures, with 3c84 and 0303+472 used as gain calibrators, and Uranus as the flux calibrator.

\subsection{IRAM}\label{NOEMA_obs}
With the goal of obtaining better resolution on the unresolved SMA detection, we observed the field around \bd~at 230.5~GHz during the winter periods in 2015-16 and 2016-17 using seven antennae with the IRAM NOEMA interferometer at Plateau de Bure at an altitude of 2552 meters.  The first session used the compact {\it BD} array configuration, with projected baseline lengths $<$\,200\,m on 2015 November 27 (2.1 hours) and  2015 December 02 (1.1 hours), the second session used the extended {\it A} array configuration, with projected baseline lengths up to 800\,m on 2017 January 03 (6.4 hours). All observations were conducted during fair weather conditions (precipitable water vapour $\sim$2-3\,mm) apart from those on 2015 December 02 which were not used in our data processing. The data were processed with the Widex correlator over a bandwidth of 3.6\,GHz centered on 230.5\,GHz in both polarizations combined to provide total intensity visibilities. Phase calibrators J0136+478 and J0300+470 were used to monitor the atmospheric phase variations once every 30 minutes to remove its effect on the target phase.  Absolute flux density calibration relies on the monitoring program of standard calibrators (MWC\,349 and LkHa\,101) conducted routinely by IRAM. At 1.3~mm, the flux density uncertainty can be up to 20\%.

\begin{deluxetable*}{lcccl}
\tabletypesize{\scriptsize}
\tablecaption{Imaging Observations for \bd}
\tablewidth{0pt}
\tablehead{ 
\colhead{Date (UT)} & 
\colhead{Observatory/Instrument} & 
\colhead{Wavelength/Mode} &
\colhead{Integration Time} & 
\colhead{Notes}
}
\startdata
\textbf{Scattered light imaging}  &                           &                                              &                                          &                                                                \\
\hspace{0.1in}2013 September 25        & WMKO/NIRC2     & 2.2 $\mu$m coronagraphy    & 50$\times$30s = 1500s    & parallactic angle rotation = 17.8$^\circ$          \\
\hspace{0.1in}2014 October 02             &      -                       & 2.2 $\mu$m coronagraphy   & 130$\times$20s = 2600s  & parallactic angle rotation = 36.0$^\circ$           \\
\hspace{0.1in}2015 January 10             &     -                        & 2.2 $\mu$m  coronagraphy  & 64$\times$24s = 1536s    & parallactic angle rotation = 20.6$^\circ$           \\ \\ \hline 
%2014 November 07      &      -                       & 2.2 $\mu$m, 3.8$\mu$m      & one hour?                         & parallactic angle rotation?  \\ %H and L-band data only (Hinkley PI) 
%2014 November 11      &      -                       & 2.2 $\mu$m  imaging            & one hour?                         & parallactic angle rotation?  \\ %Bowler. Is disk visible? 
%2015 September 25     &      -                       & 3.8 $\mu$m  imaging            & one hour?                         & parallactic angle rotation?  \\ %L-band by Justin and Erica Gonzalez. 
\textbf{1.3 mm imaging}  &                              &                                               &                                          &                                                                \\
\hspace{0.1in}Various                            & JCMT/SCUBA-2   &  850 \um~imaging             &          4.2\,hr                      &                                                                  \\ 
\hspace{0.1in}2015 August 10               & SMA                     & 224 GHz imaging                  &          6.0\,hr                      & Filler Observation          \\
\hspace{0.1in}2015 November 27         & IRAM/NOEMA      & 230.5 GHz imaging               &          2.1\,hr                      & CD configuration, resolution: 2'' \\
\hspace{0.1in}2015 December 02         &             -                &                 -                            &          1.1\,hr                      & CD configuration, resolution: 2'' \\
\hspace{0.1in}2017 January 03             &             -                &                   -                          &          6.4\,hr                      & AZ configuration,  resolution: 0.5'' 
%???                              & IGRINS                  &                                              &                                          &    
\enddata
\label{obstable}
\end{deluxetable*}

\begin{figure*}[ht]
  \centering
  \vspace{-1.5in}
  \includegraphics[angle=0,width=1.1\textwidth]{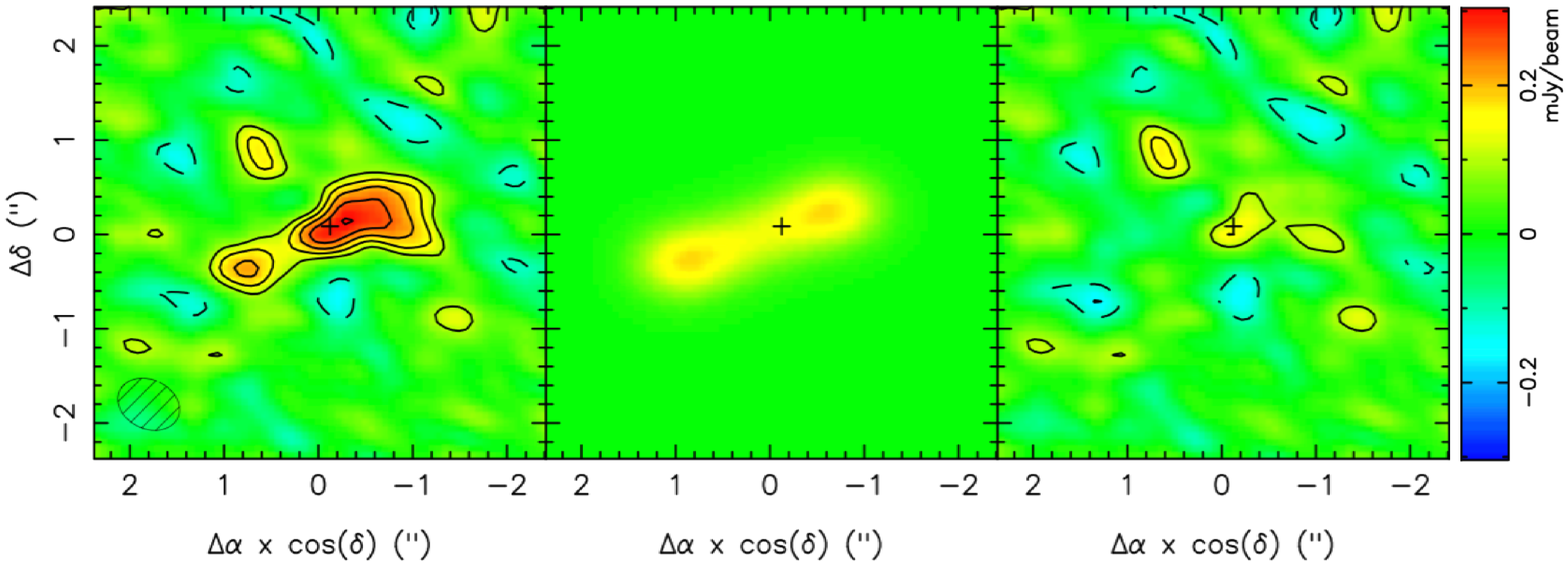}
  \vspace{-1.5in}
  \caption{  {\it Left:} NOEMA CLEAN map of the structure around \bd.  {\it Middle:} best-fit model of the disk adjusted to the map as described in \S\ref{sec:noema_mod}.  {\it Right:} residual map when the model has been subtracted from the CLEAN map.  
  The contours range from -3,-2,+2,+3...6$\sigma$ in intervals of 1$\sigma$, where $\sigma$ is 42$\mu$Jy/beam. The color coded scale at the right is in units of mJy/beam, and the beam size is 0.68$^{\prime\prime}\times 0.52^{\prime\prime}$.  The cross in the map corresponds to the position of the star given in \S\ref{NOEMA_obs}.
  }
  \label{fig:lestrade}
\end{figure*}

%\subsection{NOEMA Data Post-Processing}
The NOEMA data were processed with the IRAM software GILDAS. The data acquired at the three epochs of observations were combined to provide satisfactory uv-coverage. Measured complex visibilities were Fourier inverted and deconvolved using the CLEAN algorithm \citep{h74} to provide the final intensity map show in Figure~\ref{fig:lestrade}. The data processing used natural cleaning with an elliptical support large enough to include the source and keep some sidelobes ($5''$ by $10''$ with a position angle close to the disk orientation in the Keck image). One thousand iterations were carried out, but the convergence of the flux density was satisfactorily reached after a few hundred. The image has a beam size of ($0.68'' \times 0.52''$) and is limited in extent by the primary beam of individual antennas ($18''$ at 230~GHz). 

In the CLEAN map shown in Figure~\ref{fig:lestrade}, we find a millimeter-wavelength structure detected with a statistical significance of 4-5$\sigma$, elongated and aligned with the edge-on disk seen in the Keck image (see discussion of the NOEMA data modelling in \S\ref{sec:noema_mod}). The NOEMA structure is well centered at the optical position of \bd~provided by the \textit{Gaia} DR2 catalogue \citep{gbv18} updated with its proper motion at the observation midpoint ($\alpha=02^h21^m13.15055^s\pm0.00013^s$ and $\delta=+46^{\circ}00'06.4333''\pm0.0011''$ at 2016.0). The Gaia and NOEMA celestial reference frames are aligned at the submilliarcsecond level \citep{l20}. 

The elongated NOEMA structure in the map of Figure~\ref{fig:lestrade} is almost certainly millimeter-sized dust that traces a belt of planetesimals centered on \bd~corresponding to the Keck scattered light images. We note that the west side of the structure is more prominent than the east side, but our modelling finds that any brightness asymmetry is only at the $\sim$2$\sigma$ level.  The integrated flux density of the disk at $\lambda$=1.3\,mm is 1.11\,$\pm$\,0.17\,mJy (statistical uncertainty). This measurement was done by fitting an ellipsoid to the data in the uv-plane with a priori values conjectured from the image itself. The photospheric flux density of \bd~is three orders of magnitude smaller at this wavelength and so does not contribute. Finally, using the beam full-width at half maximum combined with the distance to \bd, we can infer that the belt width is smaller than $35$~AU. 
%(NOEMA beam FWHM $\times$ $73.4\pm0.31$~pc in the GAIA catalogue).

\subsection{IGRINS}\label{sec:igrins}
For the purpose of kinematic analysis, we obtained an improved radial velocity of the star by observing with the IGRINS spectrograph \citep{pjy14} at McDonald Observatory on 2015 October 23. 
A total exposure time of 360s was obtained in the ``ABBA'' observing pattern, and 200s were obtained for the purpose of telluric calibration on the star HD14212.  Cross-correlation against other objects with similar spectral types provides an absolute radial velocity of 0.18$\pm$0.22 km\,s$^{-1}$.

\section{Results \& Model Fitting}\label{sec:modeling}
Below we describe the result of our modelling of the 2.2\um~Keck scattered light images, as well as the IRAM interferometric imaging at 1.3mm.

\subsection{Coronagraphic $K^{\prime}$-band Scattered Light Imaging}
Our $K^{\prime}$-band coronagraphic images are presented in Figure \ref{fig:three panel} for the three epochs where we model the disk geometry.   No disk signal was detected during our 2014 November 11 observations, and so we do not model these data. 
For the three epochs shown in Figure \ref{fig:three panel}, we detect a peak signal-to-noise (SNR) ratio for the disk of 4.8, 4.8, and 4.7, respectively. The bottom panel of Figure \ref{fig:coadded} also shows a signal-to-noise (SNR) map for the combined dataset, indicating a peak SNR of 5.8. Figure \ref{fig:coadded} (top panel) also shows a measure of the disk surface brightness, indicating a peak surface brightness of $\sim$1-2\,mJy/arcsec$^2$. However, it should be noted that in the speckle dominated regime at inner working angles $\lesssim$0.5$^{\prime\prime}$, some ``self-subtraction'' due to the KLIP algorithm will diminish the surface brightness of the disk by a factor of $\sim$2. 

To constrain the disk geometry, we use an injection modelling process as described in \citet{wlb14} and \citet{mhv17}. We fit a circular disk with five free parameters to the data. These parameters are the position angle, inclination angle, disk radius, forward scattering (parametrized by the Henyey-Greenstein parameter $g$), and an overall scaling term. Synthetic model images of an optically thin ring are generated using the GRaTeR radiative transfer code \citep{alm99}, and then convolved with a Gaussian function to reproduce the resolution of the telescope.  To generate disk models with GRaTeR, we use the following parametrization. The disk density is calculated as

\begin{equation}
    \rho = \frac{\rho_{\circ} \exp  \left( \left[\frac{-|z|}{\xi_\circ}\left(\frac{r}{r_{\circ}}\right)^{-\beta}\right]^\gamma \right) }
    {\sqrt{\left(\frac{r}{r_{\circ}}\right)^{-2\alpha_\mathrm{in}}+
        \left(\frac{r}{r_{\circ}}\right)^{-2\alpha_\mathrm{out}}}},
\end{equation}

where $r$ is the radial distance from the disk center, $z$ is the vertical distance from the disk midplane, and the remaining terms are parameters defining the disk vertical and radial structure. For this work, the only disk parameters we fit are $\rho_0$ and $r_0$; values for the other parameters are fixed at $\alpha_{in} = 10$, $\alpha_{out} = -3$, $\xi_0 = 1$, $\beta = 0$ and $\gamma = 2$, 
i.e. we fit a thin, narrow ring and do not attempt to constrain the vertical disk structure since this is unresolved in the Keck/NIRC2 data. 

The scattered light contribution from each point in the disk is calculated as:
\begin{equation}
F \propto \frac{\rho \times \textrm{p}(\theta)}{d^2},
\end{equation}
where $\theta$ is the scattering angle of the dust, and $d$ is the distance from the star to each point, and $\textrm{p}(\theta)$ is the scattering phase function. In each fit, the disk is scaled by an arbitrary value $\rho_0$, which takes into account the effect of various factors including the stellar luminosity, stellar distance and telescope gain.  For $\textrm{p}(\theta)$ we use the Henyey-Greenstein scattering function:

\begin{equation}
    \textrm{p}\left(\theta\right) = \frac{1}{4\pi} \frac{1-g^2}{\left[1-2g\cos{\theta}+g^2\right]^{\frac{3}{2}}}.
\end{equation}
This calculation is performed over a grid encompassing the entire disk, and contributions along each line of sight are added to create a final disk image.   As well as the $\rho_0$, $r_0$ and $g$ disk parameters, we fit two geometrical viewing terms, namely the position angle and the inclination of the disk, denoted by ``PA'' and i$_{\rm tilt}$.

\renewcommand{\arraystretch}{1.4}
\begin{deluxetable}{llll}
\tabletypesize{\scriptsize}
\tablecaption{Best fit model parameters to our $K^{\prime}$ scattered light data. $\rho_0$ values are quoted relative to the first epoch.}
\tablewidth{0pt}
\tablehead{ 
\colhead{} & 
\colhead{2013 Sep 25} & 
\colhead{2014 Oct 02} &
\colhead{2015 Jan 10} 
}
\startdata
PA [$^\circ$]                     &  $109.9^{+0.6}_{-0.7}$      & $109.5^{+0.9}_{-1.0}$       &   $110.2\pm1.2$ \\
%Note that the PA values had 90 degrees added to the values that Elisabeth sent on 24 Aug 2020. 
i$_\textrm{tilt}$ [$^\circ$] &  $86.9^{+3.0}_{-1.6}$      & $87.1^{+3.2}_{-1.7}$       &   $86.1^{+2.2}_{-2.0}$ \\
%Note that to be consistent with the NOEMA values, the itilt values above have been calculated by 
%subtracting the values  Elisabeth sent on 24 Aug 2020 (93.1, 92.9, 93.9 degrees) from 180 degrees. 
$\rho_0$                          &  $1.0^{+0.9}_{-0.4}$        &  $0.7^{+1.0}_{-0.3}$        &  $0.9^{+0.7}_{-0.4}$  \\
r$_0$ [au]                        &  $71^{+19}_{-17}$           & $70^{+16}_{-27}$            & $68^{+20}_{-11}$ \\
g                                      &  $0.44^{+0.32}_{-0.23}$  & $0.28^{+0.43}_{-0.17}$   & $0.33^{+0.36}_{-0.16}$ 
\enddata
\label{tab:fittable}
\end{deluxetable}

These models are then rotated and subtracted from each individual data frame in the cleaned data cubes. These data cubes with negative disks injected are processed using the same KLIP algorithm discussed in \S\ref{sec:keck_observations}.
By injecting negative disk images into the cleaned data, we mimic a forward-modelling procedure whereby the throughput of the KLIP algorithm is accurately taken into account in the disk modelling. To determine the goodness-of-fit of each disk model, we calculate the $\chi^2$ of a rectangular region enclosing the disk, with central pixels close to the star excluded.

%The error on each pixel is calculated as a function of radius as the standard deviation of pixels outside this rectangular region.  
The uncertainty per pixel is calculated as a function of radius. We first take the standard deviation of all those pixels outside the mask, in small annuli centered on the star. We then inflate these errors to account for the correlation between pixels. The inflation term is the square root of the number of pixels per correlated region, and is calculated from both the FWHM of the instrument and the radial elongation of speckles due to the filter width (see Figure 1). At radius $r$, the inflated error is $\sigma(r)  = \sqrt{FWHM \times r\Delta\lambda/\lambda} \times \sigma_{pixel}(r)$, where $\lambda$ and $\Delta\lambda$ are the central wavelength and bandpass of the NIRC2 $K^{\prime}$ filter.
Using this method we find an initial fit value for the first epoch of data using a downhill minimization algorithm. We then initiate Metropolis Hastings MCMC chains from this initial value to explore the parameter space, and calculate the best fit and uncertainty values. MCMC chains are generated with \texttt{emcee} \citep{fhl13}, probabilities are assigned to each sample as $\exp {\left(-\chi^2/2\right)}$, and we use uniform priors.

We find good fits to each of the three epochs, with the parameters showing good agreement between epochs.  
%The best-fit values at each of the three epochs are presented in Table~\ref{tab:fittable}. 
Images of the three best-fit models are shown in Figure~\ref{fig:three panel} and best-fit parameters are listed in Table~\ref{tab:fittable}, where we list the median and uncertainties which include 34\% of the samples on each side of this median value. The disk is $\sim$3$^\circ$ from edge on, and has a radius of $\sim$70\,AU, although the radius is poorly constrained due to the viewing geometry. Given the relatively low signal-to-noise ratio of the disk, we are not able to detect additional structure with the current data.

\begin{figure*}
      \vspace{0.0in}
  \centering
  \includegraphics[width=0.5\textwidth]{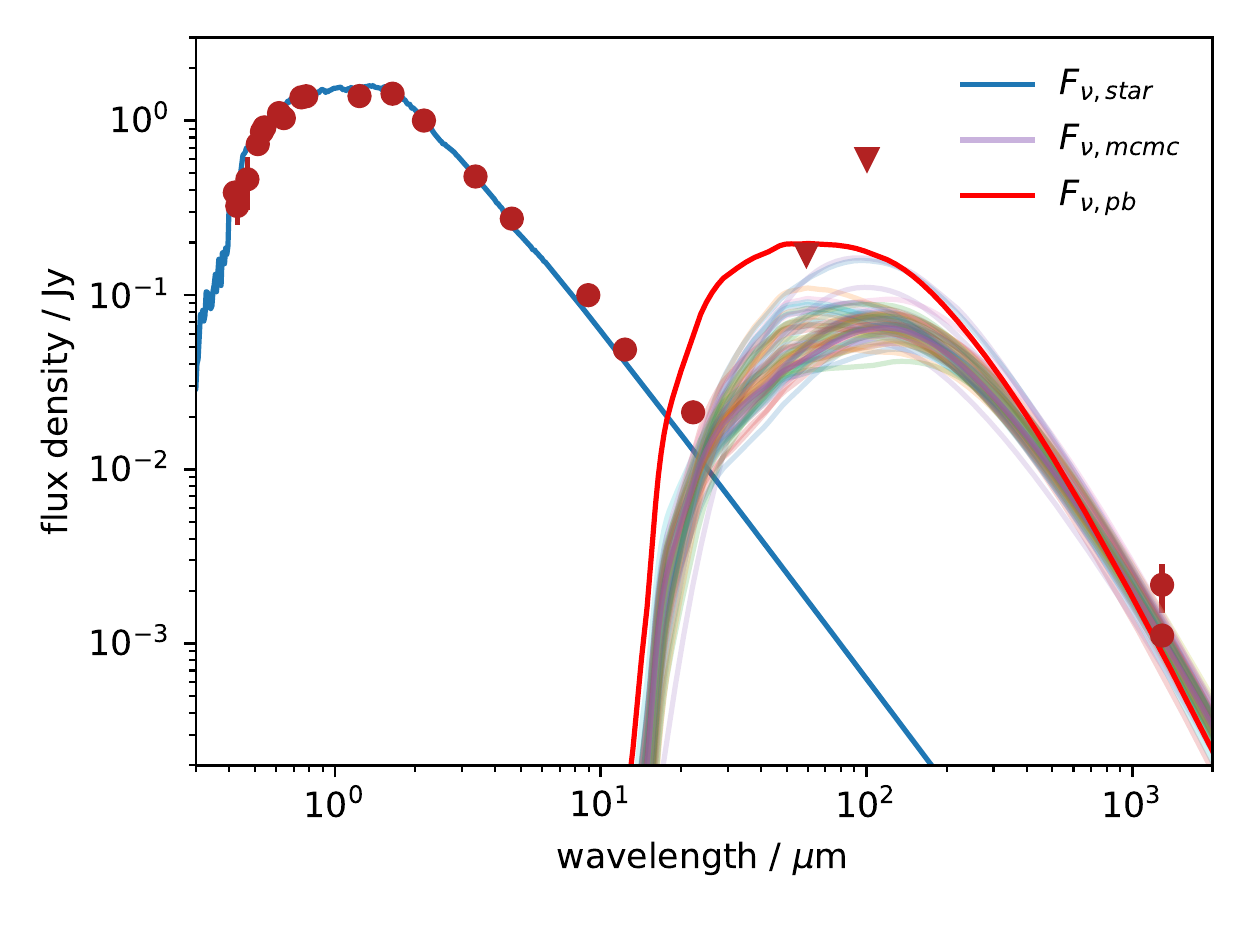},\includegraphics[width=0.5\textwidth]{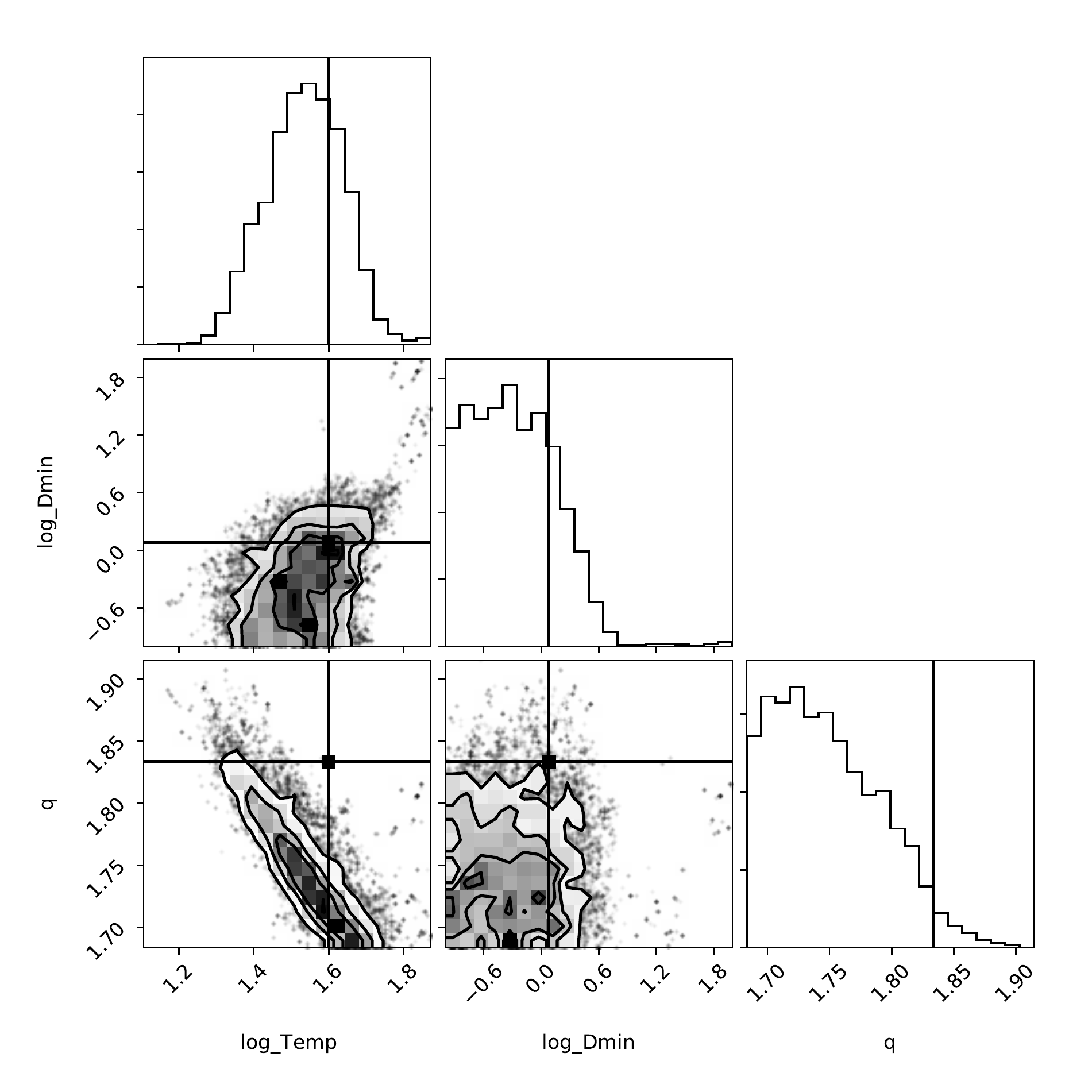} 
      \vspace{-0.25in}
  \caption{ {\it Left :}
  Flux density distribution for \bd, showing observed photometry (filled circles). The blue line denoted by $F_{\nu, star}$ shows the 5280~K stellar model, and the theoretical model of the disk parent belt based on the NOEMA flux density is shown by the red line ($F_{\nu, pb}$). The MCMC best fits of the silicate model  are shown by the semi-transparent lines.  {\it Right:} The posterior distributions of the silicate belt parameters with the MCMC chains (crosses are parameter values for the theoretically expected model). 
}
  \label{fig:SED+CMD}
\end{figure*}

\subsection{IRAM NOEMA Imaging at 1.3mm}\label{sec:noema_mod}
As the Keck scattered light data and the IRAM NOEMA 1.3mm data are sensitive to dust populations at different radii, we have chosen to model the Keck and NOEMA data separately, and not to perform a simultaneous fit.  We have  modelled the NOEMA image with the thermal emission of an optically thin ring and again use a Markov chain Monte Carlo (MCMC) approach using \texttt{emcee} \citep{fhl13}. 
The fitting is done by generating disk images, which are then convolved with the NOEMA beam, and then subtracted from the data to create a residual map (from which we compute a $\chi^2$ value). With this approach, the NOEMA radius is constrained to  ${0.85''}^{+0.18}_{-0.25}$ ($62^{+13}_{-18}$~AU) comparable to the Kuiper belt radius, the inclination is  $87^{\circ +0.46}_{~-0.45}$  and the PA is $109.85^{\circ +0.11}_{~-0.09}$. The belt radius derived from the NOEMA data is smaller than that derived from the Keck scattered light imaging ($r_0\sim$\,70\,AU in Table \ref{tab:fittable}) as expected since scattered light imaging will be sensitive to micron-sized grains that are likely blown to high eccentricity orbits by radiation pressure, while the NOEMA dataset will reveal millimeter-sized grains tracing the belt of larger planetesimals. 
Higher dynamic range maps would be needed for a detailed comparison of the NOEMA radius with the radius parameter $r_0$ of the grain surface density in the GRaTeR model used to fit the scattered light image.  Lastly, with the current data, we cannot place strong limits on the belt width besides our constraint that it must be less that 35\,AU based on FWHM argument mentioned in \S\ref{NOEMA_obs}.  

\subsection{SED Modelling of \bd} %Start new section on SED modelling 
To construct the SED shown in Figure~\ref{fig:SED+CMD}, we collected optical (Tycho-2, APASS, Gaia) and infrared (2MASS, AKARI, WISE, IRAS) photometry for \bd. We fit the SED with various models, using PHOENIX spectra for the stellar photosphere \citep{hba97}, and amorphous silicate spectra for the dust opacity \citep{lg97}. The silicate models use three parameters; a size distribution $n(D) \propto D^{2-3q}$ with a minimum size $D_{\rm min}$, a temperature that corresponds to the blackbody equilibrium temperature $T_{\rm bb}$ at the assumed stellocentric distance, and a normalisation parameter (e.g. mass, area, or fractional luminosity). These models are fitted using \texttt{multinest} \citep{fhb09} as described in \citet{yg18}, and also using Markov chain Monte Carlo (MCMC) using \texttt{emcee}.

We fit two models, both of which have the same stellar component and assume the same silicate optical properties. The first is a ``theoretical'' model, in which the dust size distribution is fixed to $q=11/6$, the minimum grain size to $D_{\rm min}=1.2$\,$\mu$m, and the dust blackbody temperature to 40\,K (as expected at 62\,AU, the best-fit parent belt location from modelling the NOEMA data). The only parameter for the dust component in this model is the normalisation, and it should fit the data if these fairly standard assumptions are correct. The second model is the same, but parameters are allowed to vary.

The best fit stellar model in both cases has an effective temperature of $5280 \pm 100$\,K, consistent with our analysis in \S\ref{sec:age}. The luminosity is $L_*=1.4L_{\sun}$, for which the blowout grain size is 1.2\um. The dust models are shown in Figure~\ref{fig:SED+CMD}; the red line shows the theoretical model given the parent belt location as imaged by NOEMA ($F_{\nu,\,{\rm pb}}$), while the semi-transparent lines show a handful of dust models from MCMC fitting that are consistent with the data ($F_{\nu,\,{\rm mcmc}}$). The posterior distributions of the silicate belt parameters are shown in Figure~\ref{fig:SED+CMD}, which also shows the parameters of the theoretically expected model. The theoretical model is inconsistent with the posterior distributions for the fitted model, whose size distribution slope is significantly shallower. This discrepancy is caused by the NOEMA flux density being higher than predicted given the mid-IR excess seen with WISE; this pushes the fitted models towards shallower size distributions, whose behaviour can be understood in terms of the models presented by \citet[][]{kw14}, e.g.~Figure 11 in that work.

The astrosilicate model used here therefore illustrates that there may be a discrepancy between the mid-IR and mm-wave data, in terms of our ability to explain both with a dust population at a single stellocentric radius. This might be resolved with deeper mm-wave imaging to better locate the belt; a more distant location (i.e. a cooler $T_{\rm bb}$) would allow a steeper size distribution to fit the data (Figure~\ref{fig:SED+CMD}). Alternatively, the dust spectrum may not come from a single location, for example comprising both warm inner dust as well as cool outer dust. Nonetheless, we use our fitted model of the infrared excess to calculate a fractional luminosity of $L_{IR}/L_{tot} =6^{+2}_{-1}\times$10$^{-4}$, higher than that of the debris disk around AU Mic \citep[$L_{IR}/L_{tot}\simeq3.5\times$10$^{-4}$;][]{mks15}.  
%We note that our fractional luminosity is somewhat different than the $L_{IR}/L_{star}$=3.9$\times$10$^{-4}$ value quoted in \citep{cs16}. However, 

\begin{figure*}[ht]
  \centering
  \includegraphics[width=0.9\textwidth]{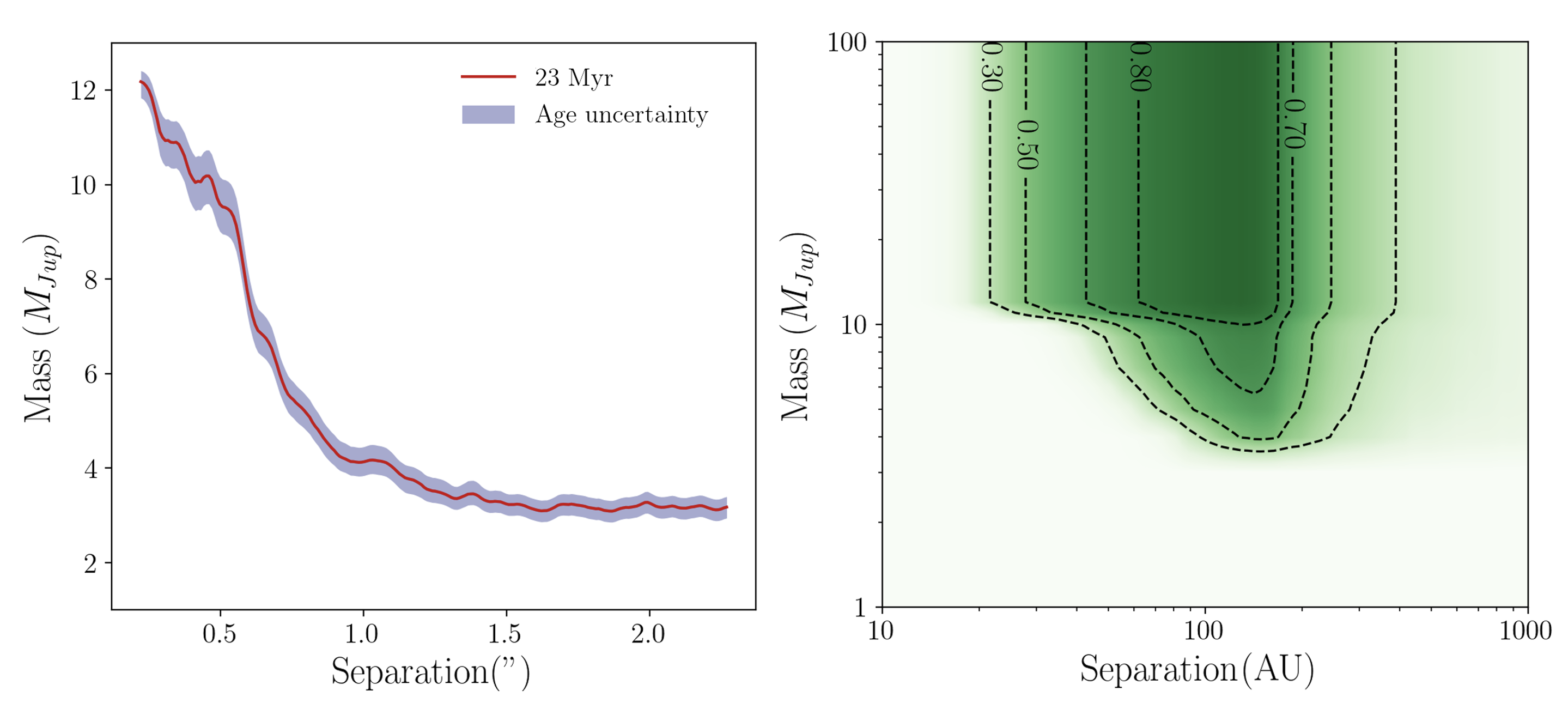}
  \caption{\textit{Left:} Our measured $K^{\prime}$-band scattered light contrast limit expressed as the lowest achievable mass (in Jupiter masses) versus orbital separation in AU using the evolutionary models of \cite{ptb20}. To bound our contrast limit, we also incorporate the upper and lower estimates of the age of the BPMG (27 and 21 Myr, respectively). \textit{Right:} a probability map for the detection of planetary mass companions calculated using \texttt{Exo-DMC} \citep{b20} using the best estimate age of 23 Myr. 
  }
  \label{fig:contrast_curve}
\end{figure*}

\subsection{Sensitivity to Low Mass Companions}\label{sec:contrast_curve}
\label{sect:contrast}
No significant evidence for point sources that may be self-luminous, massive companions to \bd~was found within our Keck scattered light images. To assess the sensitivity of our near-infrared coronagraphic observations to substellar companions in the vicinity of \bd, we start with the angular measure of our achieved 5$\sigma$ contrast relative to the host star (i.e. a ``contrast curve'').  Here we present the contrast performance for only the 2013 September 25 epoch, as this dataset delivered the best constraints on planetary mass companions that might be dynamically interacting with the disk. We then use this measure of our 5$\sigma$ contrast, together with the models of \citet{ptb20} assuming an age of 23$\pm3$\,Myr for the BPMG, to calculate the lower mass limit of substellar objects that would be detectable with our observations described in \S\ref{sec:keck_observations}. Figure~\ref{fig:contrast_curve} shows this sensitivity to substellar companions as a function of projected orbital separation from the host star. Figure~\ref{fig:contrast_curve} also shows a detection probability map (right panel)  based on our estimated contrast curve, and generated using the Exo-DMC algorithm \citep{b20}, demonstrating that our observations are sensitive to $\sim$8\,M$_{Jup}$ objects at 50 AU, and $\sim$3\,M$_{Jup}$ objects at $\sim$100 AU.  In the left panel of Figure~\ref{fig:contrast_curve}, we also indicate the upper and lower estimates of the age of the BPMG (27 and 21 Myr, respectively), however, these different ages do not substantially change our overall sensitivity.

\section{Age of \bd}\label{sec:age}
\bd~was first characterized in a spectroscopic survey of ROSAT All-Sky Survey (RASS) X-ray sources by \citet{gkf09}. Of the $\sim$1000 ROSAT sources studied in that work, \bd~had the third highest equivalent width of Li\,I $\lambda$6707.8 (EW $\simeq$ 300.8\,$\pm$\,11.6 m\AA), which is above the upper envelope of values seen for Pleiades members. Independent of any other considerations, this observation constrains the star's age to empirically be $\lesssim$100 Myr \citep[e.g.][]{lsm05}. \citet{ks13} report that the star exhibits saturated X-ray emission, \loglxlbol = -3.36\,$\pm$\,0.14, and its ASAS photometry shows that the star is a relatively fast rotator (4.73 day period) with strong $V$-band variability (amplitude $\simeq$ 0.07 mag). 

\begin{figure*}
  \centering
  \includegraphics[width=1.0\textwidth]{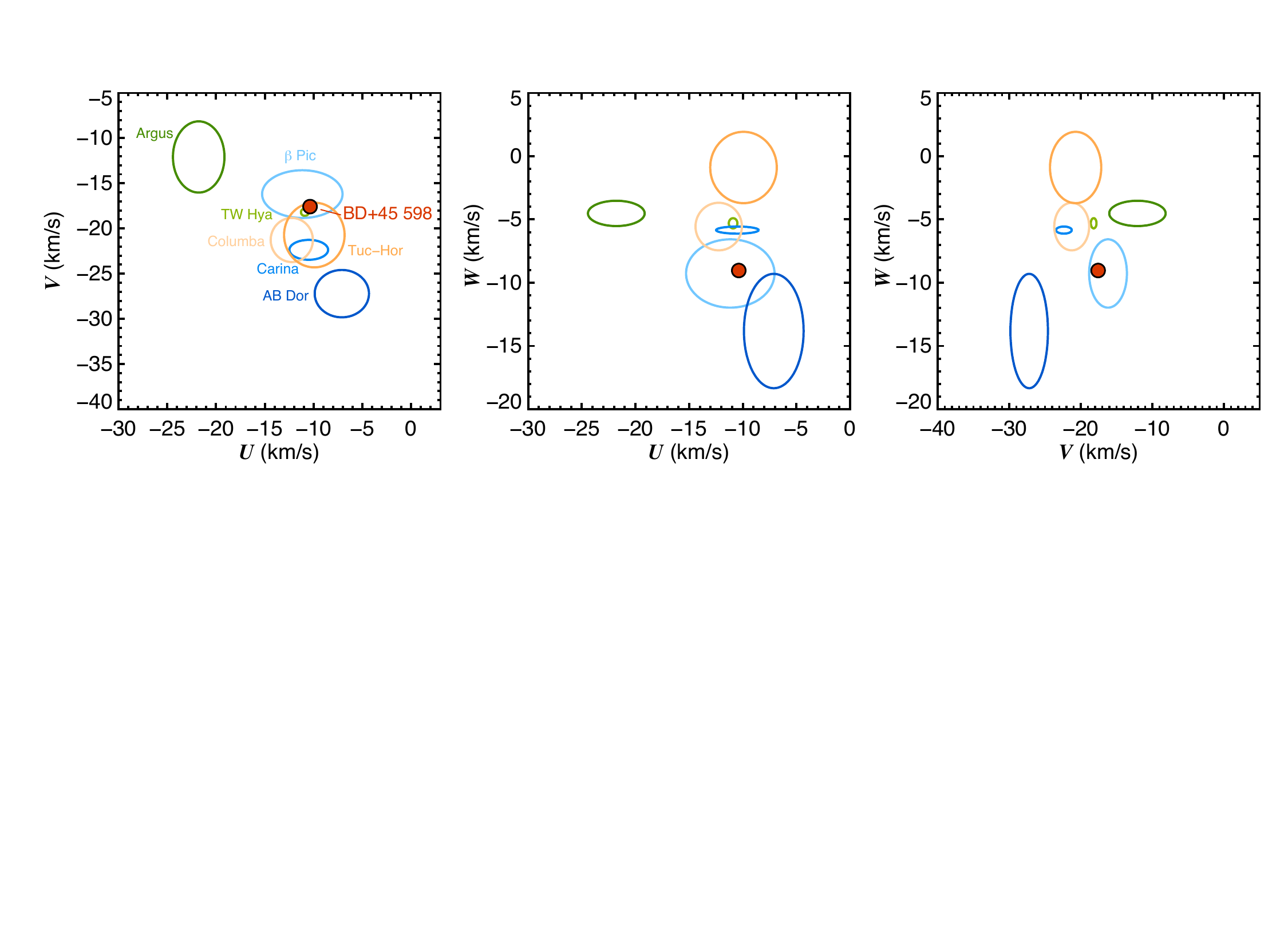}
    \vspace{-2.5in}
\caption{The galactic space velocity representation for several nearby young moving groups (ellipses), as defined from \citet{mdf14}, except for TW Hya taken from \citet{wab13}. The position of \bd~in this space is shown by the red point, and in all three projections is consistent  with membership to the BPMG.}
  \label{fig:uvwplot}
\label{uvwplot}
\end{figure*}

\citet{mpk11} claimed that \bd\, is comoving with the star HD 15745, another star with a debris disk imaged in scattered light \citep{kdf07} situated 9$^{\circ}$ away, and they further argue based on kinematics and youth indicators that both stars are likely to be members of the BPMG. The median age of the BPMG is now constrained to $<$15\%\, accuracy \citep[23\,$\pm$\,3 Myr, ][]{mb14, bmn15}. \citet{Binks20} used the \citet{GaiaDR2} astrometry and an adopted radial velocity of \vrad\, = -1.191\,$\pm$\,1.045	km\,s$^{-1}$ to estimate a Galactic velocity for \bd~of $U, V, W$ = -10.52, -17.32, -9.04 ($\pm$0.77, $\pm$0.67, $\pm$0.26) km\,s$^{-1}$. Using the Bayesian classification tool BANYAN $\Sigma$, as well as the velocity and position centroids of \cite{gmm18}, these authors also remark that \bd~is closest to that of BPMG, the 32 Ori group, and the $\epsilon$ Cha group. \citet{Binks20} notes that the membership probability is highest (yet still meager: 0.3\%) for BPMG, providing further support for the membership assignment of \citet{mpk11}. We adopt the new Gaia EDR3 astrometry \citep{GaiaEDR3} and an updated average radial velocity\footnote{Unweighted mean of reported radial velocities from \citet{gkf09}, \citet{mpk11}, \citet{GaiaDR2}, and this study.} of \vrad\, = -0.8\,$\pm$\,0.3 km\,s$^{-1}$, and estimate an updated velocity of $U, V, W$ = -10.8, -17.0, -9.2 ($\pm$0.2, $\pm$0.2, $\pm$0.2) km\,s$^{-1}$. In Figure~\ref{fig:uvwplot} we show the $U, V, W$ galactic space velocity for \bd~compared with the representations of several young moving groups taken from \citet{mdf14} and \citet{wab13}, showing that in all three projections, the Galactic space velocity of \bd~is consistent with membership to the BPMG. However, running the revised astrometry and radial velocity through the BANYAN $\Sigma$ kinematic membership tool of \citet{gmm18}
results in an updated membership probability of 0.2\%\, to BPMG and 99.8\%\, to "field" (with no other groups breaking above 0.1\%).

The reasons for the relatively low probability of membership to the BPMG returned by BANYAN $\Sigma$ for \bd~are clear, and primarily stem from its galactic position. Current models of the spatial/kinematic distribution of the BPMG have been primarily constructed using stars within $\sim$20-50\,pc \citep[e.g.][]{zsb01}.  Therefore, when presented with a target like \bd~at $\sim$73\,pc, tools such as BANYAN $\Sigma$, will automatically assign a lower membership probability. Indeed, more distant stars like \bd~that also show multiple clear signs of youth may be essential for extending the model of BPMG members to greater distances. Additionally, the star lies only $\sim$14$^\circ$ above the galactic plane and has relatively small proper motion, which substantially increases the chances of it being assigned as an interloper to any association by such tools. However, the numerous pieces of evidence pointing to the youth of this star discussed above essentially rule out any chances of being an interloper. 

The measured parallax of \bd~from Gaia means that a HR diagram analysis can be performed to analyze the potential membership to the BPMG. When we compare the $G-K_s$ versus $M_G$ color-magnitude data for the BPMG members in the survey by \citet{Binks20}, we find that \bd\, ($G-K_s=1.68$, $M_G=4.40$) 
appears right on the single-star trend for BPMG members (approximately $M_G$ = 1.13 + 1.968($G-K_s$), for 1.5 $<$ $G-K_s$ $<$ 4.0). 
For the $G-K_s$ color of \bd, one would expect a typical BPMG member to have $M_G$ $\simeq$ 4.44, remarkably within 0.04 mag of that observed for the star. 
Thus, we agree with the assessment of \citet{mpk11}, finding that the proper motion of \bd, radial velocity, color-magnitude diagram position, presence of lithium, and X-ray emission, {\it all} seem to be consistent with membership of the star to the BPMG.

%A comparison of the $BVJHK_s$ colors for \bd~with the dwarf color sequence of \citet{pm13} shows that \bd~has an optical/near-IR SED consistent with an unreddened dwarf with T$_{\rm eff}$ $\simeq$ 5240\,K.  This is consistent with the spectroscopic T$_{\rm eff}$ estimate of 5149\,$\pm$\,99K from \citet{gkf09}.  Using its trigonometric parallax distance, \bd~would have absolute magnitudes of M$_V$ = 4.73 and M$_{Ks}$ = 2.72 (uncertainties $\sim$0.16 mag). A comparison of the combination of $V-J$ color (1.44) and M$_V$ with the locus of color-magnitude diagram points for bona fide BPMG members shows that the star would be indistinguishable from other group members. Similarly, when

%Despite these numerous lines of evidence pointing to membership in the BPMG, we find that the BANYAN $\Sigma$ tool \citep{gmm18}, a Bayesian classification tool for determining the membership probability of stars to nearby young associations, returns a significant probability ($>$99\%) that this star is \textit{not} a member of any moving group or young association. 

\section{Summary and Conclusions}\label{sec:conclusions}
We have identified a nearly edge-on circumstellar debris disk, imaged in scattered light at the W.M. Keck Observatory, associated with the nearby K1 star \bd.  We have shown that this star's proper motion, radial velocity, color-magnitude diagram position, presence of lithium, and X-ray emission, are all consistent with membership in the BPMG with age 23\,$\pm$\,3 Myr.  Using the NOEMA interferometer at the Plateau de Bure Observatory operating at 1.3\,mm, we also find a 4-5$\sigma$ detection of a source aligned with the orientation of the disk structure in the 2.2\um~images.  Our best fitting model to the disk scattered light images is one with a peak disk radius at $\sim$70\,AU, and a fractional luminosity $L_{IR}/L_{tot}\simeq6^{+2}_{-1}\times$10$^{-4}$. 
%consistent with our {\color{red} best fit model of the system's spectral energy distribution}. 
Recently \cite{mpa20} and \cite{tpb20} presented the discovery of CP-72 2713, another BPMG member with a cold debris disk, and a relatively high fractional luminosity of 1.1$\times 10^{-3}$, suggesting it is a more massive, dust-rich ``sibling'' to the debris disk system associated with AU Mic \citep[$L_{IR}/L_{tot}\simeq3.5\times$10$^{-4}$;][]{mks15}.  With a fractional luminosity of $L_{IR}/L_{tot}\simeq6^{+2}_{-1}\times$10$^{-4}$, our study suggests \bd~could also be considered another dust-rich counterpart to AU Mic. However, some differences between AU Mic and \bd~are apparent. Namely, previous studies of AU Mic using the SMA and ALMA find a central radius for the emission belt of 35-40AU \citep{wam12,mwr13}, while our observations find a significantly larger radius of $\sim$70\,AU. 

Systems like \bd~and CP-72 2713 may be examples of newly identified late-type members of YMGs that provide an opportunity to study the evolution of debris disks starting at the earliest ages. Discovery of more such systems would complement previous infrared and millimeter surveys for debris disks within nearby young moving groups \citep[e.g][]{rsw08, drc12, rbm14, mka16, hmk17, bj17}.  While the youngest associations of stars have been efficient at producing scattered light systems \citep[e.g. Sco-Cen,][]{ekf20}, debris structures in nearby YMGs span a range of ages from $\sim$10-100\,Myr, which will allow us to image in sequence the evolution of circumstellar environments after the era of massive planet formation has ceased.  
Thus, with its young age of $\sim$20\,Myr, and reasonably close proximity ($\sim$70\,pc), \bd~will be an excellent target for upcoming high contrast imaging missions in the northern hemisphere \citep[e.g.][]{cbd18}, to search for massive companions at close orbital separations ($\sim$5-20 AU), or possibly even much lower-mass companions at wider orbital separations (20-200 AU) using the upcoming James Webb Space Telescope \citep{hcb18,chb21}.

\acknowledgments
We thank the anonymous referee for several helpful comments, which improved the quality of the manuscript.  
Some of this work was performed in part under an NSF Astronomy and Astrophysics Postdoctoral Fellowship under award AST-1203023.  
Some of the data presented herein were obtained at the W.M. Keck Observatory, which is operated as a scientific partnership among the California Institute of Technology, the University of California and NASA. The Observatory was made possible by the generous financial support of the W.M. Keck Foundation.  The authors wish to recognize and acknowledge the very significant cultural role and reverence that the summit of Mauna Kea has always had within the indigenous Hawaiian community.  
Part of this work has been carried out within the framework of the National Centre of Competence in Research PlanetS supported by the Swiss National Science Foundation. ECM acknowledges the financial support of the SNSF.
GMK is supported by the Royal Society as a Royal Society University Research Fellow.  
The Submillimeter Array is a joint project between the Smithsonian Astrophysical Observatory and the Academia Sinica Institute of Astronomy and Astrophysics and is funded by the Smithsonian Institution and the Academia Sinica.  
This work used the Immersion Grating Infrared Spectrometer (IGRINS) that was developed under a collaboration between the University of Texas at Austin and the Korea Astronomy and Space Science Institute (KASI) with the financial support of the Mt. Cuba Astronomical Foundation, of the US National Science Foundation under grants AST-1229522 and AST-1702267, of the McDonald Observatory of the University of Texas at Austin, of the Korean GMT Project of KASI, and Gemini Observatory. 
This paper includes data taken at The McDonald Observatory of The University of Texas at Austin.
Part of this research was carried out at the Jet Propulsion
Laboratory, California Institute of Technology, under a contract with
the National Aeronautics and Space Administration (80NM0018D0004).

%% curly braces. If you forget this, LaTeX will generate the error
%% different from previous examples. The natbib system solves a host

\bibliography{MasterBiblio_Sasha_short_authors}
\bibliographystyle{apj.bst}

\end{document}